\documentclass[12pt]{article}
\usepackage{graphicx}
\usepackage{amssymb,amsmath,amsfonts,palatino,amsthm}
\usepackage{amssymb}
\usepackage{epstopdf}
\newcommand{\beq}{\begin{equation}}
\newcommand{\eeq}{\end{equation}}

\newcommand{\f}{\begin{equation}}
\newcommand{\ff}{\end{equation}}

\begin{document}

\title{A perspective on the landscape problem \\}
\author{Lee Smolin\thanks{lsmolin@perimeterinstitute.ca} 
\\
\\
Perimeter Institute for Theoretical Physics,\\
31 Caroline Street North, Waterloo, Ontario N2J 2Y5, Canada}
\date{\today}
\maketitle

\begin{abstract}

I discuss the historical roots of the landscape problem and propose criteria for its successful resolution. This provides a perspective to evaluate the possibility to solve it in several of the speculative cosmological scenarios under study including eternal inflation, cosmological natural selection and cyclic cosmologies.

Invited contribution for  a special issue of Foundations of Physics titled {\it Forty Years Of String Theory: Reflecting On the Foundations.}
\end{abstract}

\newpage

\tableofcontents

\section{Introduction: what is the landscape problem?}

The landscape problem\cite{LOTC} in string theory was born in 1986 when Andrew Strominger discovered that there were vast numbers of consistent string vacua-many more than the Calabi-Yao solutions which had been discovered the year before\cite{Andyland}.  As Strominger wrote in his paper describing that discovery,

\begin{quotation}
{\it
The class of supersymmetric superstring compactifications has been enormously enlarged. . . . It does not seem likely that [these] solutions . . . can be classified in the foreseeable future. As the constraints on [these] solutions are relatively weak, it does seem likely that a number of phenomenologically acceptable . . . ones can be found. . . . While this is quite reassuring, in some sense life has been made too easy. All predictive power seems to have been lost.

All of this points to the overwhelming need to find a dynamical principle for determining [which theory describes nature], which now appears more imperative than ever\cite{Andyland} .}

\end{quotation}

However compelling this now seems, it took 17 years, till 2003, for the landscape problem to be widely appreciated\cite{KKLT,lennyland}.
Why was the existence of the landscape so hard to accept?  

Part of the answer is that many of us believed in the possibility of a principled explanation for the laws of nature.  We hoped to discover a short list of principles, which could be realized in a unique theory, which would retrodict the standard model and uniquely predict the physics to be discovered beyond it.  It was reasonable to guess that these principles would be the conjunction of those of quantum theory and general relativity, given the difficulty of making any theory that combined them.  The shocking implication of the results Strominger  reported in 1986 was that this was not to be, at least within the confines of string theory. Given the implications, it is not surprising if it  took until 2003 for many researchers to give up the notion of a uniquely predictive string theory\cite{KKLT,lennyland}.  
 
String theory offered more, however, then just a vast or infinite list of possible low energy effective theories.  It offered the promise of a setting in which the different perturbative string theories are realized as expansions around solutions of a still more fundamental theory.  Within that theory there could be non-perturbative processes that would result in dynamical transitions between the different perturbative string theories; hence between different low energy phenomenologies.  That more fundamental theory would have to be background independent-by definition-because it would have to be defined in a way that transcended perturbative expansions around particular semiclassical solutions.  

Unfortunately, so far that promise of a truly background independent formulation of string theory has not been achieved\footnote{For several reasons this cannot be the $AdS/CFT$ correspondence.  One reason is that the cosmological constant has the wrong sign, another is that a cosmological theory cannot have boundaries or asymptotic regions, for reasons I will discuss below.}.  The reasons this project-often called the search for $\cal M$ theory-has so far frustrated would seem to be the subject or another review.  But the landscape problem and the problem of background independence are closely linked.  The former is the only route the latter has to experimental confirmation.  It is also the case that the background independent theory provides the setting for dynamical processes by which the universe can evolve through sequences of effective theories.  

Hence, the options for resolving the landscape problem are going to influence what we expect from the solution to the background independence problem.  Below I will suggest that this influence takes us to surprising places, altering fundamentally our strategy and expectations to construct the background independent theory.  To get there we begin by reviewing 
 the landscape problem.  Our aim will be to  put it in a broader setting which will give us a perspective from which to discuss the status of attempts to resolve both the landscape problem and the problem of finding a background independent formulation of string or $\cal M$ theory.  

The hypothesis underlying all approaches to the landscape is that there is a cosmological setting in which different regions or epochs of the universe can have different effective laws.  This implies the existence of spacetime regions not directly observable, because there is good evidence against the variation of the parameters of the standard model on scales of billions of years.  These regions must either be in the past of our big bang, or far enough away from us to be causally unrelated.  These different possibilities give rise to three classes of cosmological settings in which to situate hypotheses for the dynamical evolution of effective laws.  

\begin{enumerate}

\item{} Causally distinct regions, which may be called pluralistic multiverse cosmologies.  The main example of these is eternal inflation\cite{Vilenkin-eternali,Linde-eternali}.  

\item{} A succession of past epochs, which are cyclic cosmological models.  Recent examples are the ekpyrotic 
models of Steinhardt, Turok and colleagues\cite{PaulNeil} and the conformal cyclic cosmology of Penrose\cite{ccc}.

\item{} Combinations of the two, in which there are epochs in the past which branch on reproduction.  Examples of these are 
phoenix cyclic cosmologies\cite{phoenix}   and cosmological natural selection\cite{CNS,CNS-2, me-alt-AP,mestatus, LOTC}.

\end{enumerate}

Attempts to solve the landscape problem have so far been situated mainly within two of these settings, which are eternal inflation and cosmological natural selection.  One of the aims of this essay is to give an evaluation of how well each attempt is doing.  Another is to provide a deeper perspective that might guide the search for new solutions to the landscape problem\footnote{This partly reflects conclusions reached in joint work with Roberto Mangabeira Unger\cite{withroberto}.}.

But, before we discuss details it is worth stating carefully what exactly would constitute a solution to the landscape problem. 

\begin{itemize}

\item{}An acceptable solution to the landscape problem would be an explanation for the choice of effective laws and their parameters observed in our universe which
is scientific in the sense that

\begin{enumerate}

\item{}That explanation has several further necessary consequences at least some of which are falsifiable\cite{popper}  by presently doable observations of experiments.  

\item{} It will have other consequences which are strongly confirmable by presently doable observations or experiments.  By strongly confirmable I mean that the consequences are sufficiently unique to this explanation that if seen it would constitute strong evidence for the correctness of the explanation.

\item{}Some of the falsifiable consequences have failed attempts at falsification and some of the strongly confirmable consequences have been confirmed.   

\end{enumerate}

\item{}The solution should explain the improbable features of the standard model which include the large hierarchies in scales and dimensionless parameters and the fact that they seem to be fine tuned to create a universe that has highly improbable complex structures over a wide range of scales from clusters of galaxies down to molecular biology.

\item{}The solution should explain the very special initial conditions of the universe as well.  Given that a solution to the landscape problem requires a speculative cosmological scenario, if confirmed there will not be an opportunity to go back and adjust it to explain also the choice of cosmological initial conditions.  

\end{itemize}

Note that if a proposal to resolve the landscape problem does not satisfy conditions one and two, it cannot be considered to be even a candidate for a solution to the landscape problem. 

The landscape problem represents a serious crisis in the development of science.  Its solution requires, as we shall see,  the construction of speculative cosmological scenarios, which posit regions or epochs of our universe for which we presently have no observable evidence.  Nonetheless we must insist on taking seriously only scenarios and hypotheses that make falsifiable or strongly verifiable predictions, otherwise people can just make stuff up and the distinction between science and mythology becomes porous.  To do so is very challenging. This is hard, serious science, and it cannot be rushed.  

As challenging as this situation is, there are two reasons for optimism. One is that there already are candidate solutions that make real, falsifiable predictions.  Some of these predictions involve features of the CMB that code information about a cosmological epoch to the past of our big bang.  This 
possibility of making observations which test hypotheses about  past epochs is the second reason for optimism-and I will discuss examples  below.  

So there is no need to despair.  Meanwhile, special pleading that the standards of science should be lessoned to admit explanations with no falsifiable consequences, in order to keep alive a bold speculative idea, should be strongly resisted.  While speculation has its place in science, ultimately science is not interested in what might be true, it is interested only in what can be convincingly demonstrated by deductions from observational evidence.  

I should stress that the landscape problem, while it arose in string theory, is likely to be there whatever the fundamental unification of physics turns out to be.  There is no evidence from any approach to quantum gravity that mathematical consistency or the existence of a low energy limit restricts the matter content of a theory.  In loop quantum gravity and spin foam models it appears that  the theory is consistent with coupling to a large set of gauge groups, fermions and scalar fields.  

But the strongest reason to expect the landscape problem is not an anomaly of string theory is that it has deep historical roots, which I sketch in the next section.  It might have been anticipated a long time ago-and indeed it was. These historical roots of the landscape problem suggest that the landscape problem was bound to occur as physics progressed.  As I will argue, it is an inevitable consequence of the general form we have assumed for physical theories since Newtonian mechanics.   Thus, whether string theory is correct or not as an hypothesis about unification of the fundamental forces, the landscape problem is likely to be a feature of whatever correct theory replaces it. Therefor the future of theoretical cosmology will to a large extent hinge on finding the correct scientific resolution of the landscape problem.  

After the historical sketch I explain in chapter 3 the three main options for cosmological settings within which a solution to the landscape problem might be situated. These options are evaluated in section 4.

\section{Historical roots of the landscape problem}

As our knowledge of the elementary particles and fundamental interactions grew dramatically during the 20th Century we began to be interested in questions that are not answered by knowing the laws of physics.  One of these is the question:  {\it why are these the laws, rather than other possible laws?}   If there are many possible laws, each as logically consistent as those we observe, what selected the set that are realized in our universe?   Another question not answered by knowing the laws is what selected the initial conditions, at or near the big bang.   These questions bothered a few physicists and philosophers long before Strominger uncovered the landscape problem within string theory.  

Within living memory, the idea of the evolution of laws was energetically championed by John Archibald Wheeler, who was, uniquely,  a pioneer of both nuclear physics and the quantum theory of gravity.  In the 1960's Wheeler contemplated the bouncing of black hole\footnote{which he named.} singularities to new universes.  By a bounce he meant that quantum effects would eliminate the singularities of classical general relativity and lead regions of spactime tending to future singularities to expand again, thus forming new regions of spacetime to the future of where those singularities would have been\footnote{The evidence that quantum gravity effects eliminate singularities in this way has become much stronger recently. Compare older papers on bounces which employ mainly semiclassical methods\cite{bounce-old} to the newer literature on loop quantum cosmology which shows that bounces are generic in exact quantum evolutions of a class of quantum cosmological models\cite{LQC-bounce}. A study of modifications of coupling constants during bounces is in \cite{jr-bounce}.}.   He further hypothesized that the laws of physics-or at least 
their parameters- were ``reprocessed" on each such instance\cite{wheeler}

A generation earlier, Dirac had proposed that laws of physics may evolve,

\begin{quotation}

{\it At the beginning of time the laws of Nature were probably very different from what they are now.  Thus, we should consider the laws of Nature as continually changing with the eoch, instead of as holding uniformly throught space-time\cite{dirac}.}

\end{quotation}

Why were these great scientists drawn to speculate that laws evolve?  The reason is that absent a principled explanation for the laws we find, the evolution of laws is a necessary part of any explanation of the  why these laws problem.  There are in science only two ways to explain why some state of affairs has come about.  Either there are logical reasons it has to be that way, or there are historical causes, which acted over time to bring things to the present state.  When logical implication is insufficient, the explanation must be found in causal processes acting over time.  This was understood clearly more than a century ago by Charles Sanders Pierce, the founder of the school of philosophy called American pragmatism,

\begin{quotation}

{\it  To suppose universal laws of nature capable of being apprehended by the mind and yet having no reason for their special forms, but standing inexplicable and 	irrational, is hardly a justifiable position. Uniformities are precisely the sort of facts that need to be accounted for. Law is par excellence the thing that wants a reason. 	Now the only possible way of accounting for the laws of nature, and for uniformity 	in general, is to suppose them results of evolution\cite{CSP}.}

\end{quotation}

Pierce is insisting that it is not enough to know the laws of nature.  He is demanding that there be explicable reasons for the laws of nature themselves.  
This demand has a long precedence in the history of science.  It was most clearly articulated by Leibniz who enunciated his\cite{leibniz}

\begin{itemize}

\item{}{\bf Principle of Sufficient Reason.}  {\it For every property of nature which might be otherwise, there must be a rational reason which is sufficient to explain that choice.}  

\end{itemize}

Leibniz makes it clear the kind of sufficient reason he has in mind must be something beyond mathematical consistency.

\begin{quotation}

{\it The great foundation of mathematics is the principle of contradiction or of identity, that is to say, that a statement cannot be both true and false at the same time and that A is A, and cannot be not A.  And this single principle is enough to prove the whole of arithmetic  and the whole of geometry, that is to say all mathematical principles.  But in order to proceed from mathematics to physics another principle is necessary.  As I have observed in my Theodicity, that is, the principle of a sufficient reason, that nothing happens without there being a reason why it should be thus rather than otherwise\cite{Leibniz-psfr}.}

\end{quotation}

If no reason can  be given the choice must be a false choice.  For example, no rational reason can be given for why is the universe where it is and not ten feet to the left. From this Leibniz draws the conclusion that space must be relational, so that only relative positions within the universe are physically meaningful.  Physicists use the principle of sufficient reason in this negative way when we show that two gauge equivalent configurations of the electromagnetic or gravitational field refer to the same physical state.  Otherwise, one would not have unique deterministic evolution of fields.  
This was the essence of Einstein's "hole" argument that led him to conclude that diffeomorphism invariance is a gauge symmetry and it was also the basis of Dirac's influential work on gauge invariance in constrained Hamiltonian systems.   

What can count as sufficient reason for laws of nature?  The landscape issue is precisely the fact that mathematical consistency alone cannot account for the choice of the laws we observe governing phenomena in our universe.

Pierce is saying that if we demand sufficient reason for the choice of the laws of nature we can only answer successfully by positing that the present laws are the result of evolution from a past when the laws were different.   To put Pierce's argument in one line,  {\it Laws must evolve to be explained.}

\subsection{Sufficient reason for cosmological initial conditions}

We can also apply Leibniz's principle of sufficient reason to the problem of the selection of the initial conditions of the universe.  It is a fact that in general relativity-and presumably in any field theory of gravitation-there are an infinite number of solutions of the field equations which have an initial singularity.  To apply general relativity to cosmology, it is then necessary to give the initial conditions at-or shortly after-the singularity.  The choice of initial conditions requires explanation.  If we are optimistic and believe all questions about the universe are answerable, then  that explanation must satisfy the principle of 
sufficient reason. 

If no sufficient reason can be given within a given theory, then that theory must be wrong.  It is one thing for general relativity to have an infinite number of solutions in the asymptotically flat case, for these correspond to idealized, approximate descriptions of subsystems of the universe.  These come in many copies, so the theory must have many different solutions.  But why should a cosmological theory have an infinite number of solutions when there is only a single universe?  Why does general relativity so extravagantly overperform its job, giving not just predictions for the actual universe but also predictions for an infinite number of universes that never exist?  The only conclusion to draw is that general relativity is not the correct cosmological theory. It is-at the very least- to be supplemented, either by a theory of initial conditions or by an historical explanation which explains why such special initial conditions were picked out for realization in the one real world.

\subsection{The common roots of the landscape problem and  background independence}

I bring up Leibniz's principle of sufficient reason because I believe it helps greatly to clarify the issues that we confront in seeking a solution to the landscape problem.  In particular, the principle of sufficient reason not only tells us that {\it laws must evolve to be explained} and that there must be a dynamical explanation for the initial conditions of the universe.  It greatly constrains the context in which we seek those dynamical explanations which led to the selection of the laws and initial conditions of our universe.  This is because the principle of sufficient reason singles out a class of theories of space and time-those that are relational and background independent.  

Thus, the landscape problem and the problem of making a background independent quantum theory of gravity and cosmology are profoundly linked.  Conceptually and historically, both have their roots in the Principle of Sufficient Reason. 

This is seen in several corollaries of the principle of sufficient reason that are highly relevant for the contemporary search for an explanation for the selection of the observed laws of physics.

\begin{itemize}

\item{}{\bf Space and time are relational rather than absolute.}  Leibniz used the principle of sufficient reason heavily in his debates against Newton's concept of absolute space and time.  These arguments were elaborated by Mach and became the motivation for Einstein's invention of general relativity.

\item{}{\bf The principle of the identity of the indiscernible.}  This holds that there cannot be two entities in the world with exactly the same properties. If all properties are relational then to identify a particular elementary particle is the same as giving a list of all its properties, including its relative position in space and time.  

\item{}{ \bf Causal or explanatory closure.}  This is the demand that all chains of causes and chains of explanations for events in the universe close within the universe. 

\item{}{\bf A cosmological theory must be spatially compact, without boundaries.}  This is a consequence of the previous principle, because if one has to impose asymptotic boundary conditions to get a solution to a field equation then chains of causation point outside the universe.   The only way to avoid having to input boundary conditions to determine a solution to general relativity or any other field theory is to not have any spatial or null boundaries.  This was the reason Einstein introduced spatial compact solutions for general relativity and insisted on their use in modeling cosmology.  

\item{}{\bf No unreciprocated actions.}  There should be no entity which acts on dynamical degrees of freedom, which is not itself a dynamical degree of freedom which is acted back on in return.  This principle was used by Einstein to motivate his rejection of absolute space in favour of a theory-general relativity-in which the geometry of spacetime is a dynamical degree of freedom.  Another way to say this is that there must be no absolute or ideal entities, whose properties influence the evolution of degrees of freedom, which are themselves not dynamically determined.  

\item{}{\bf Background independence.}  As a consequence of the last principle, we must rule out proposals for a quantum theory of gravity which are dependent for their formulation on fixed, non-dynamical, classical geometries.  This rules out as fundamental perturbative formulations of quantum gravity and string theory.  These must be approximations to background independent formulations in which geometry is fully dynamical and fully quantum.   

\end{itemize}

The demand for an explanation for the choices of laws and initial conditions is deeply related to the requirement that theories of gravitation and spacetime be background independent.  Both demands reflect the need to explain the universe only in terms of dynamical processes internal to it.  Both reflect the need that a scientific explanation for the laws and initial conditions of our universe not rest on conjectures that are beyond experimental check.  There must be no fixed background geometry for space and time, for the same reason there can be no fixed, timeless laws.  Either would leave us with a description of the universe failing the  test of sufficient reason.  

This connection is one link between the landscape problem and the problem of making a background independent formulation of string or $\cal M$ theory. 

\subsection{What kinds of explanation count as sufficient reason in science?}

What kind of explanations can count as sufficient reason for a law or theory?   As I argued before, two general kinds  
 of explanations that could be advanced to account for a state of affairs.  Reasons can be logical or they can be historical.  
 They both may serve, but they have very different consequences for the methodology of science.  This is because  
 logical explanations can be complete while our knowledge of the past is always incomplete.  This raises the question of whether we can ever give the complete sufficient reason for a feature of the present world whose explanation has an historical component.  
 
 Even if our knowledge of past conditions is detailed enough to justify giving a sufficient reason for a present fact in terms of conditions at some past time, there remains the problem that a complete explanation would involve us explaining those past conditions.  So causal explanation can often involve us in a regress in which we just push the mysteries deeper into the past.  

It might be thus be objected that sufficient reason is too strong of a criteria to apply to explanations in science.  It rarely happens that we know enough about nature to give the complete reason something has a given property rather than an alternative. Explanation in science is almost always tentative, for two distinct reasons.  Scientific explanations are always subject to revision when knowledge advances.  These revisions can involve the replacement of theories by newer theories which explain the same evidence in different terms.  As technology and science advance we also learn more about the past.

How are we to reconcile the demand for sufficient reason with the tentative character of explanation in physics? 

We do so by acknowledging that our attempts to give sufficient reason for present features of the universe, including the laws, may in many cases be incomplete.  But we can still be faithful to the demand for sufficient reason by accepting explanations that leave room for further developments that may improve our understanding of the past.  There is no shame in explaining the plentitude of galaxies in the universe in terms of postulated initial conditions shortly after the initial singularity, so long as we leave open the task of explaining those initial conditions. This is the case if we postulate that the initial singularity postdicted by classical general relativity was in reality a bounce from an earlier epoch. By doing so we invite the possibility of explaining those initial conditions as being the result of evolution through the bounce.  

It is, on the other hand, against the spirit of the demand for sufficient reason to adopt explanations for present conditions that close off further inquiry.  
We do this if we adopt an apriori principle which posits that the inferred initial conditions are the result of a unique initial state of the universe.  The problem with this may be that the principle is not subject to falsification or further verification because it is alleged to only act once-at the beginning of the universe.  An apriori principle which is designed to postdic the initial conditions inferred from experiment without making any additional predictions can not be falsified.  The adoption of this kind of explanation stops further inquiry and so is against the spirit of the demand for sufficient reason.  

This concern is addressed by the principles of causal closure and no unreciprocated actions.  If all chains of explanation remain within the universe and involve only dynamical degrees of freedom that have reciprocal interactions with everything they influence, then our explanations of the present will go only so far as our knowledge of the past.  

We can call this the {\it practice of modesty about past causes}: do not assert an untestable hypothesis that closes off further investigation where a confession of ignorance leaves the situation open to genuine explanation when our knowledge of the past improves.

To summarize, in some circumstances, the demand for sufficient reason must result in a confession of ignorance, when causal chains are pushed back into the past to the point where our present knowledge of the past ends.  This is better then proclaiming first movers or initial states which are not subject to further explanation in terms of their pasts, and so cannot be further improved as our observations of the past improve.  


We can illustrate this principle of modesty with one of the main choices that face contemporary cosmologists, which is whether to accept the initial cosmological singularity as a first moment of time or hypothesize that in the correct physics the singularity will be replaced by a bounce that opens up a much older past to scientific investigation.  By endowing the very early universe with a past, the hypothesis of a bounce makes the unusual conditions of the early universe explicable in terms of its prior history. 

If one instead hypothesizes that the initial singularity the beginning of time one is face to face with the necessary of basing science on an inexplicable choice, which is the initial conditions of the universe.  

\subsection{The full scope of the landscape problem}

To emphasize the full import and scope of the landscape problem, we have to dig a bit deeper.  As I will now argue,  its roots lie in assumptions physicists have made since the time of Newton about the structure of physical theories.  Newtonian mechanics, field theory, general relativity and quantum theory have a common framework which is organized as follows.

\begin{enumerate}

\item{} We specify a system to be studied. This is almost always an approximate or effective description of a small subsystem of the universe.  We do this by specifying the degrees of freedom we are interested in studying. 

\item{} We specify a timeless space of states, $\cal C$, which is a phase space or Hilbert space.  This gives us possible initial states in which our system can be prepared.

\item{} We specify a dynamical law acting on $\cal C$. This allows us to evolve states in $\cal C$, so that given the choice of initial state,  prepared at an initial time, we can compute the state at any future time. 

\end{enumerate}

We call this framework of theories, the {\it Newtonian paradigm}.   

Because this framework has been so successful when applied to the small subsystems of the universe, it appears almost obvious that when we come to the task of developing a cosmological theory, we should just scale it up to include the whole universe in the state space, $\cal C$.  However, 
as successful as it has been, this schema for physical theories cannot be applied to the universe as a whole.  There are several distinct reasons for this.

\begin{itemize}

\item{} The Newtonian paradigm takes the dynamical law as input. It cannot be the basis of explaining why that law is the one that applies to our universe.  Hence to adopt this paradigm as the framework of a cosmological theory will leave a great deal of mystery in our understanding of the world.  We will fail to fulfill the demand to give sufficient reason for every cosmological question.

\item{} Similarly, the Newtonian paradigm takes the specification of the initial state as input. It cannot justify or explain the choice of initial state.  The demand for sufficient reason will again not be answered.

\item{} The Newtonian paradigm assumes that there is an absolute distinction between the role of state and the role of the dynamical law.  This distinction can be operationally realized on small subsystems because we can prepare a system many times in different initial states and observe what aspects of the resulting evolution are universal and what are consequences of the choice of initial state.  The dynamical law is inferred from observations of universal features of the motion which are independent of the choice of initial state.  When we come to the universe there is only a single history and so we have no way to operationally or experimentally distinguish the role of the law from the choice of initial state.  

This can be a practical as well as a theoretical issue because there can be degeneracies in cosmological models arising from the fact that a single observation can be explained equally well by modifying the law of motion or the choice of initial conditions.  One sees examples of this in attempts to fit inflationary models to data such as the possible non-guassianities\cite{nongauss}. 

\item{}  Any theory formulated in the Newtonian paradigm will have an infinite number of solutions.  But,  the universe is unique-so only one cosmological history is physically real.  The Newtonian paradigm is then very extravagant when applied to cosmology because it not only makes predictions about the future of the one real universe, it offers predictions for an infinite number of universes which are never realized.  The Newtonian paradigm cannot explain why the one solution that is realized is picked out from the infinite number of possibilities.  

\end{itemize} 

Because of these issues, it is best to think of attempts to construct an exact cosmological theory by scaling up the Newtonian paradigm as fallacious-because they take a method that is very well suited to describing subsystems of the universe and apply it beyond its realm of validity.  This can be called the {\it cosmological fallacy.}

Note that this does not apply to commonly studied cosmological models such as the standard cosmological model as these are explicitly based on truncations to a  subset of the degrees of freedom, namely the homogeneous degrees of freedom and small fluctuations around them in a limited region.  

One way to express the cosmological fallacy is through the following  {\it  cosmological dilemma.}  The Newtonian paradigm expresses the forms of all the laws we know which have been thought of as exact.  Nonetheless, {\bf  every law formulated and verified within the Newtonian paradigm can only apply to a bounded domain and hence is approximate.}

Here is the argument for this perhaps counterintuitive assertion.  To operationally realize the distinction between laws and initial conditions one needs many instances to apply the law to. Only that way can one show that the law holds when the initial conditions are varied.
But since there are many instances, each one is only a part of the history of the universe. Hence each instance is a description of a subsystem of the universe. However, there are no true isolated systems in nature. Each treatment of an isolated system is an approximate truncation of an open system.  Hence each such theory is approximate.

Only a truly cosmological theory could be an exact theory.  But our different arguments tell us that we can only hope for sufficient reason within a cosmological theory that does not fall into the Newtonian paradigm.  

If one ignores the cosmological dilemma and proceeds to try to construct a theory of the whole universe within the Newtonian paradigm one is committing the cosmological fallacy.

The landscape issue is then an aspect of the failure of the Newtonian paradigm to serve as a basis for a truly cosmological theory.  If we are to solve the landscape problem and not fall into an infinite regress, we must do so in a new framework which cannot be characterized as within the Newtonian paradigm.  The task is then to invent a new theory which can be applied to the universe as a whole, which will not leave us asking, why this theory and why these initial conditions?

How can we do this?  By seeking theories which transcend the three absolute features of the Newtonian paradigm: that the space of states is timeless, that the choice of laws is timeless and that there is an absolute distinction between laws and states.  

These comments apply to any theory expected to hold at a cosmological scale: hence they apply to the background independent framework from which, it is hoped, the different perturbative string theories will emerge.  Hence we learn something important about the search for $\cal M$ theory: it must not be describable within the Newtonian paradigm.  If it is to be the real theory of everything it must be formulated in such a way that we cannot further enquire as to why this law and why these initial conditions.  It must somehow furnish its own sufficient reason.

\section{Options for a solution to the landscape problem}

Yet another way to state the landscape problem is to assert that our observable universe does not contain enough information to answer the two big why questions:  Why these laws and not others? and Why those initial conditions?   But if we insist on the principle of explanatory closure than the universe must contain enough information to answer any query that can be made about its properties-including these two questions.  The answer must lie in regions of the universe that we have not so far observed directly.  For it is likely that the universe is bigger and older than the region we can observe.  

In our analysis of the why these laws question we concluded that laws must have evolved dynamically to be explained.  This implies that there were dynamical processes in our past by which the laws evolved.  As we do not see any evidence that the fundamental laws or their parameters evolved in the observable past, these processes must have gone on in regions yet unresolved observationally.  This accords with the intuitive picture that the effective laws may have evolved in events that involved energies or energy densities much in excess of those in our observable universe.

We now face several choices:

\begin{enumerate}

\item{}Was there a bounce or singularity to our past?

\item{}Did the evolution of laws happen all at once, or incrementally over many stages?  That is, do we live in a first generation universe or does our universe have a long chain of ancestry?

\item{}Was the chain of ancestry linear, so that each universe gives rise to a single progeny, or does it branch, with each universe giving rise to many progeny.  

\end{enumerate}

Let us investigate the different options.  These give rise to three classes of global cosmological scenarios.

\subsection{Three options for global structure of the larger universe}

If we posit that the initial singularity was really the first moment of time, then there is a brief time available for the evolution of the laws to have taken place.  In this case there is unlikely to have been time for incremental evolution through many epochs.  Our universe then probably arose from some primordial state in one or a few steps.

One early suggestion for such a cosmological setting for variation of the laws was  Villenkin and Linde's eternal inflation scenario\cite{Vilenkin-eternali,Linde-eternali}, within which an infinite number of universes are born as bubbles in phase transitions from a primordial eternally inflating medium.    In the simplest version of this framework our observable universe is one of an infinite number of universes each produced in a single step from a primordial state of eternal inflation.  (It is also possible that there are bubbles within bubbles but these chains of descent are not taken as central to the explanatory power of the scenario as they are in cosmological natural selection\footnote{There can even be tunneling back to the initial false vacuum leading to a recylcing of the universe as described in \cite{recycling}.}.)

The resulting multiverse scenario posits that the infinite numbers of universes are mostly causally disjoint from each other.  A bubble universe may have collided with other bubbles, but almost certainly any pair of bubbles in the population of universes are causally disjoint\footnote{An observer in a bubble will see a finite number of bubbles colliding with theirs in the past. Eventually, given infinite time, a bubble may collide with an infinite number of other bubbles but this will still be an infinitesimal fraction of the infinite colllection of bubbles\cite{infinitecollisions} }.  We can call this a {\it pluralistic cosmological scenario.}

Although eternal inflation was proposed before the realization of the string landscape, it has become the setting in which much research on dynamical evolution of laws on the landscape has been carried out\cite{KKLT,lennyland}.  

On the other hand, by positing that the singularity was replaced by a bounce we endow our universe with a deep past during which there may have been many epochs of classical universes.  These would have allowed the effective laws to evolve incrementally over many generations, all in our causal past.  These may be called {\it cosmological scenarios with succession.}  

There are again two choices, depending on what bounced.  The big bang may have followed a complete collapse of a prior universe.   So we arrive to the scenario of a {\it cyclic universe. }

The big crunch of a cyclic universe may have given rise to a single progeny-or it may have given rise to many.  The latter may be the case if there is a selection effect whereby regions of the crunch must be sufficiently homogeneous to bounce.  Hence we have to distinguish between {\it linear cyclic cosmologies}, in which a universe has a single progeny,  and {\it branching cyclic cosmologies}, in which there will be many. 

The other possibility was that the  big bang was the result of the bounce of a black hole singularity.  If black hole singularities bounce then a universe may have many progeny, each the result of a collapse to a black hole.  Indeed, our universe can be estimated to have at least $10^{18} $ black holes and hence at least as many progeny.  Hence scenarios in which black hole singularities bounce are branching cosmological scenarios.  

The scenario of bouncing black hole singularities is the setting for the framework of cosmological natural 
selection\cite{CNS,CNS-2, me-alt-AP,mestatus, LOTC}, which will be discussed below.

The only kinds of singularities which are generic in solutions to the Einstein equations are cosmological and black hole singularities.  So these are the only options for cosmological scenarios in which singularities are replaced by bounces.  

So we have the following options for a global cosmological model\footnote{There are also hybrids of these scenarios such as \cite{hybrids}}:

\begin{itemize}

\item{}{\bf Pluralistic scenarios} such as eternal inflation in which there is a population of universes, all derived from a primordial state by a one stage process, largely if not completely causally distinct from each other.  

\item{}{\bf Linear cyclic scenarios} in which there is a succession of universes, each with a single parent and a single ancestor.

\item{}{\bf Branching scenarios} in which each universe has a single parent but many progeny.

\end{itemize}

We now investigate the options for explaining the selection of laws in our universe in each of these three kinds of scenario. 

\section{Prospects for a solution of the landscape problem in  the three scenarios}

Before we analyze the possible solutions to the landscape problem offered by the three kinds of scenarios we should be mindful of a few key issues.

\begin{itemize}

\item{} In any landscape scenario-whether in biology or physics- there are two landscapes: the landscape of fundamental parameters and the landscape of parameters of effective low energy theories.  There can be a rather complicated relationship between them.   In biology these are the spaces of genotypes-the actual DNA sequences and the space of phenotypes-the space of actual features of creatures that natural selection acts on.  In physics these may be the landscape of string theories and the landscape of parameters of the standard model.   In biology, as well as in physics, the explanatory power of a scenario depend partly on how well understood are the relationships between the two kinds of landscapes.  

\item{}The bounces are very high energy processes but there is evidence for a lot of fine tuning at the level of the low energy parameters.  How can the bounces then play a role in selecting for fine tuning of the low energy parameters?   

\item{}We can observe only what is in our past light cone.  If a cosmological scenario posits an ensemble of universes outside of causal contact with our own then we risk a situation where the characterization of the other members of the ensemble is free from check by observation.  There is a great danger then of just making stuff up to get answers we want.  The only way to constrain an ensemble of causally disconnected universes by observation is if there is a dynamical principle that makes it possible to deduce that every or almost every universe in the ensemble shares some property $\cal P$.  Then an observation that $\cal P$ is not seen would falsify the theory.  

\end{itemize}

Mindful of these cautions, we can now examine what opportunities our three kinds of cosmological scenarios offer for a solution of the landscape problem. 

\subsection{Linear cyclic models}

The linear cyclic models have a great advantage over the other two scenarios in that all the epochs or universes it posits are in the causal past of ours.  There is then abundant opportunities for making predictions that are subject to observational check.  So far two kinds of cyclic models have been studied, and both offer falsifiable predictions.  The ekpyrotic models of Steinhardt, Turok and collaborators predict that there will be no observable tensor modes in the $CMB$.   The conformal cyclic cosmology of Penrose predicts the existence of concentric circles in the $CMB$ due to gravitational waves formed by colliding black holes in the previous era\footnote{Claims by Penrose and Gurzadyan\cite{roger-circles} that these have been observed are presently controversial\cite{nocircles}.}.

What prospects then do the linear cyclic models have to explaining the selection of laws?  One can easily hypothesize that at each bounce there are changes in the effective laws, perhaps brought about by phase transitions amongst vacua of string theory or whatever the fundamental theory is.  This will give us a series of points in the landscape, representing the effective laws in each epoch.    However, {\it to explain the choice of laws there must be an attractor in the landscape.}  Otherwise the progression of laws through the epochs will just be random, and nothing about the present choice of laws will be explicable.

For the evolution on the landscape to converge to an attractor, the changes in each generation must be small.  Also, 
to explain the choice of parameters of the low energy theory by a series of transitions in the fundamental theory, it must be that small changes in the fundamental landscape give rise to small changes in the landscape of the low energy effective theory. 

Furthermore, that attractor must somehow be determined by properties of low energy physics, otherwise the fine tunings of the standard model will not be explicable.

\subsection{Branching models}

Branching models give share one good property with linear cyclic models, which is that there are long chains of descent.  This can make possible incremental accumulation of good properties through slow, stable, evolution to attractors.  However they deviate from linear cyclic models in giving rise to a growing population of causally disconnected universes.  These can lead to predictions about our universe only to the extent that it can be predicted that there will be properties, $\cal P$ shared by all or almost all members of the ensemble. 

This is illustrated by the two examples we have of branching models.

\subsubsection{Branching cyclic cosmologies}

In the branching cyclic cosmologies it can be hypothesized that only regions of the collapsing universe that are sufficiently spatially homogeneous will bounce to make new expanding universes\cite{phoenix}.  Because the region must be very homogenous to bounce, each new universe will be very homogeneous.  Homogeneity is then a property $\cal P$ that is shared by all members of the ensemble-hence it is predicted for our universe.  One can hope that more detailed modeling of the bounces may lead to new predictions for our universe of this kind which may be falsifiable.  

We can note that a great advantage of cyclic cosmologies in general is that they eliminate the need for inflation to explain the specialness of the cosmological initial conditions.  

Can the branching cyclic cosmologies explain the selection of the low energy physics?  In this regard the answer is the same as with regard to the linear branching cosmologies: the changes in both the fundamental and effective laws must be small from generation to generation and there must be an attractor in the landscape of the low energy theory for the evolution to converge to.  

\subsubsection{Cosmological natural selection}

Cosmological natural selection\cite{CNS,CNS-2, me-alt-AP,mestatus, LOTC}  was invented to give an answer to the landscape problem that explained the reasons for the fine tunings of the standard model without making use of the anthropic principle\footnote{Critiques of cosmological natural selection were published in \cite{critiques}. These have been all answered in the recent papers \cite{ me-alt-AP,mestatus} and book \cite{LOTC} (see especially the appendix and end notes.)}.     The idea was to invent a cosmological scenario that naturally explained why the universe is fine tuned for complex structures such as long lived stars, spiral galaxies and organic molecules-using the same mechanism that biology uses to generate improbable complex structure.  

This suggested that there would be in cosmology an analogue of biological fitness-the number of progeny of a universe as a function of its low energy parameters  This analogy inspired the suggestion that there would be an evolution of effective field theories on a landscape of parameters analogous to the fitness landscapes studied by population biologists\footnote{Indeed, the use of the word landscape was meant to suggest the analogy to the fitness landscapes of evolutionary biology\cite{LOTC}.}. 

This was inspired by an analogy between selection of effective laws in a cosmological setting and natural selection in a biological setting.  The theory is based on two hypotheses:

\begin{itemize}

\item{}Universes reproduce when black hole singularities bounce to become new regions of spacetime.

\item{}During the bounce, the excursions through a violent interlude at the Planck scale induces {\it small} random changes in the parameters of the effective field theories that govern physics before and after the transition.  

\end{itemize}

The analogue of biological fitness is then the average number of black holes produced in a universe, seen as a function of the parameters of the standard models of physics and cosmology.  We can call this function on the landscape the cosmological fitness. Combinations of parameters that are local maxima of this fitness function are attractors on the landscape.  After many generations the population of universes becomes clumped in the regions near these local maxima.  

The great advantage of cosmological natural selection over linear cyclic cosmologies is then that it creates attractors on the landscape.  

This comes about because the effective laws which are most common in the ensemble of universes are those that reproduce the most, which means they have the most black holes.  Thus, a property $\cal P$ shared by almost all members of the ensemble will be, after many generations, the following: {\it small changes in the parameters of the effective landscape will almost always lead to universes which produce fewer black holes.}  Another way to say this is the following: If we define the fitness of a point in the landscape by the average number of black holes produced by a universe with those parameters, then after many generations almost every member of the ensemble will be near a local maximum of the fitness.  

This explains the specialness of the tunings of the parameters of the standard model, because it turns out that several aspects of those tunings enhance the production of black holes.  These include,

\begin{enumerate}

\item{} The large ratios required for the existence of long lived stable stars, including $\frac{m_{proton}}{m_{planck}}$, 
$\frac{m_{e}}{m_{proton}}$ and  $\frac{m_\nu}{m_{e}}$, 

\item{} The coincidences among the proton-neutron mass difference, electron and pion masses, making nuclear fusion possible, as well as the sign of the proton-neutron mass difference. 

\item{} The strength of the weak interaction which appears fine tuned both for nucleosynthesis and for supernovas to inject energy into the interstellar medium, catalyzing the production of massive stars whose remnants include black holes.

\item{}The fine tunings which result in the stability and plentiful production of carbon and oxygen.  These appear to be necessary to cool the giant molecular clouds from which form the massive stars which are the progenitors of black holes as well as to provide insulation to keep the clouds cold.

\end{enumerate}

It should be emphasized that cosmological natural selection is the only of our scenarios that explains the fine tunings of the parameters of the standard model.  It does so because the cosmological scenario makes low energy physics causative of structure on a vast scale-that of the population of universes.  It does so  by strongly influencing the distribution of parameters in that population.  

This feature could be mimicked by the branching cyclic models, but only if there were some reason why having something like our present low energy physics could lead a universe to have more regions which were sufficiently homogeneous to bounce.  This is unlikely because the conditions in the final crunch are not going to be sensitive to details of the choices of parameters of low energy physics.  What cosmological natural selection accomplishes, apparently uniquely, is to make the population of universes delicately sensitive to the parameters of low energy physics.  It does this naturally and necessarily, because it takes delicate tunings of parameters to produce a large number of black  holes.  

Because of this coupling between cosmology and low energy physics, cosmological natural selection makes a few predictions that are vulnerable to falsification by present observations.   It is instructive to review three of them.

To maximize the number of black holes produced,  the upper mass limit (UML) for stable neutron stars should be as low as possible.  As pointed out by Brown and collaborators \cite{bethe-brown}, the UML would be lower if neutron stars contain kaon condensates in their cores.  That is, 
\f
UML_{kaon} < UML_{conventional}
\ff

This requires that the kaon mass, and hence the strange quark mass be sufficiently low.  Since none of the other physics leading to black hole production is sensitive to the strange quark mass (within the relevant range) cosmological natural selection then implies that the strange quark mass has been tuned so that neutron stars have kaon condensate cores.   

Both the theoretical understanding of the nuclear physics of kaon condenstate stars and the observational situation has evolved since this prediction was published in 1992\cite{LP}. 

Bethe and Brown\cite{bethe-brown} argued that a kaon condenstate neutron star would have an $UML_{kaon} \approx 1.6 M_{solar}$, so that is the figure I used initially.  However, as emphasized recently by Lattimer and Prakash, there is actually a range of predictions for $UML_{kaon}$.  These depend on assumptions about the equation of state and range upwards to two solar masses\cite{LP}.   So in the light of current knowledge the correct prediction is
\f
UML_{kaon} <  2 M_{solar}
\ff  
The present experimental situation is summarized in \cite{LP}.  There is an observation of a neutron star with a mass of $1.97$ solar masses, to good accuracy.  This is just inside the range consistent with the prediction that neutron stars have lowered upper mass limits due to having kaon condensate cores.  However, there are observations of neutron stars with wider error bars of around $2.4$ solar masses.  This, if confirmed, would be inconsistent with the prediction of cosmological natural selection. 

So while it is disappointing that the observation of a $1.97$ solar mass neutron star cannot be taken as a falsification of cosmological natural selection, that theory remains highly vulnerable to falsification in the near future.  

One question often raised is why cosmological natural selection is not ruled out by the possibility of changing a cosmological parameter to greatly increase the production of primordial black holes.  This could be done by turning up the scale of the density fluctuations, 
$\delta = \frac{\delta \rho}{\rho}$ which has been measured to be around $10^{-5}$.  

An answer can be given in the context of single field-single parameter inflation. In that theory $\delta$ is determined by $\lambda$ the strength of the self-coupling of the inflaton field.  This controls the slope of the inflaton potential and hence the number of efoldings grows with decreasing $lambda$
as $N \approx \lambda^{-\frac{1}{2}}$.  This means that the volume of the universe-and hence the number of ordinary black holes produced, scales as
\f
V \approx e^{N} \approx  e^{\delta^{-\frac{1}{2}}}
\ff
Hence, there is a competition between raising the number of primordial black  holes while exponentially shrinking the universe and so decreasing the number of black holes produced by stellar evolution.  The exponential dominates and the result is that cosmological natural selection predicts the smallest possible $\delta$ consistent with galaxy formation\cite{CNS,CNS-2}.  One makes more black holes overall by having an exponentially bigger universe and making them later from stars than one does by having a lot of primordial black holes in a tiny universe.

However this argument only works in the simplest model of inflation.  In more complex models with more fields and parameters, $\delta$ is 
uncoupled to $N$ and one can have a large universe whose black hole production is dominated by primordial black holes.  Hence, cosmological natural selection predicts that inflation, if true, must be single field single inflation whose potential is governed by a single parameter.  This is so far consistent with all observations, but it could be falsified by future observations, for example if high levels of non-gaussianity are confirmed. 

Once $\delta$ is fixed in this way, cosmological natural selection makes a prediction for the value of the cosmological constant.  This is because, if $\delta$ is small, as is observed in our universe, there is a critical value of the cosmological constant, $\Lambda_0$ such that for $\Lambda > \Lambda_0$ the universe would expand too fast for galaxies to form.  But without galaxies there would not be many massive stars which are the path way to most black holes in our universe.  Hence cosmological natural selection predicts $\delta$ small and $\Lambda < \Lambda_0$.  However the fitness function will not depend strongly on $\Lambda$ within that range, hence one can expect\footnote{This prediction was pointed to in \cite{LOTC}, which was published in 1997,  just before the discovery of dark energy. "This means that we may expect that when all the observations have been sorted out,  there will be a small cosmological constant, there will be a neutrino mass {\it and} Omega will not be exactly equal to one."\cite{LOTC}, page 315. }  that in a typical universe $\Lambda \approx \Lambda_0$.   

Note finally that the choice of initial conditions is not so far explained by the scenario of cosmological natural selection.  This is challenging as new universes arise from black hole singularities which are generically very inhomogeneous.  Thus, cosmological natural selection probably requires inflation to make sense of the specialness of the initial conditions.  

\subsection{Pluralistic cosmological scenarios}

Let us finally turn to the pluralistic scenarios, of which eternal inflation is the main example.  In this scenario an infinite population of universes is produced in one step from the formation of bubbles in an eternally inflating primordial phase.  At least in its simplest form, this lacks the strengths of either the cyclic or the branching scenarios.  

While a few other bubbles may have collided with our universe-giving a chance to confirm but not falsify predictions of the scenario\cite{bubbles}-almost all the universes in the population are causally disconnected from our own.  It is usually assumed that the universes that are created randomly sample the points in the landscape of the fundamental theory so there are almost no properties common to all universes.  The only property $\cal P$ put forward as satisfied in all universes is that the curvature should be slightly negative.  However, this will be difficult to confirm or falsify with near future observations because it will require a great deal of precision to distinguish this from vanishing curvature.  


Moreover, the formation of bubbles takes place at very high energies, typically grand unified scales where the details of the parameters of low energy physics are not going to matter.  So there is no mechanism for a coupling between the fine tunings of low energy physics and the dynamics that produces the ensemble.  in its absence, it  has to be concluded that universes like ours with fine tunings of low energy parameters are very rare. 

In the absence of any large set of properties $\cal P$, common to the ensemble proponents of eternal inflation have to 
fall back on the anthropic principle\cite{AP}.  
This has so far not led to any genuine predictions, and it is pretty clear why this is unlikely.  The properties of a universe can be divided into two classes.  The first class consists of properties that play a role in making a universe friendly to life.  Examples include the values of the fine structure constant and the proton-neutron mass difference.  Class two consists of properties that do not strongly influence the bio-friendlyness of a universe.  These include the masses of the second and third generation fermions (so long as they stay sufficiently heavier than the first generation.)

The first class of properties must hold and their verification does not provide evidence for any cosmological scenario-because we already know the universe is bio-friendly.  That is to be explained, by an argument that is not circular, ie does not assume our existence. The second class are assumed to be randomly distributed in the ensemble-hence, since they are uncoupled to bio-friendlyness they will be randomly distributed in the ensemble of bio-friendly universes.  Hence no prediction can be made for them.  

These kinds of arguments, developed in more detail\cite{LOTC,TTWP,me-alt-AP,mestatus}, make it very unlikely that the anthropic principle can ever be the basis for a prediction by which a cosmological scenario could be falsified or strongly verified\footnote{An excellent historical and critical survey of the anthropic principle and related developments is in \cite{kragh}.}.

What are we to make then of the claims that there have been successful predictions made based on the anthropic principle?  In fact, such claims must be fallacious, and they have been shown to be.  This is discussed in detail in \cite{LOTC,TTWP,me-alt-AP} but I can mention quickly here that there are basically two kinds of fallacies in these claims.  First, a statement that X is essential for life is added to an already correct argument involving X.  

For example, Hoyle argued successfully that if carbon is produced in stars there must be a certain energy level in a nuclei.  He based this successful argument on the observation that carbon is abundent in the universe.  The fact that carbon plays a role in life plays absolutely no role in the argument.

Later Weinberg\cite{weinberg}  argued that if there were to be an ensemble of universes with random values of the cosmological constant, that $\Lambda$ would be seen to have a value within an order or two of magnitude below a critical value, $\Lambda_0$ above which no galaxies form.   This had nothing to do with life, as galaxies are observed to be plentiful.  It is true that the observed value came to within five percent of the $\Lambda_0$ Weinberg used.

However, that estimate for $\Lambda_0$ depended on an assumption about the ensemble of universes which is that $\Lambda$ is the only constant to be varied.  There is no justification for this assumption. If other parameters are allowed to vary, the estimate of $\Lambda_0$ greatly increases, making the prediction far less successful.  For example if $\Lambda$ and $\delta $ are allowed to both vary the chances are quite small that their  values are both as 
small as observed\cite{varymore}.

The point is not that an ensemble with only $\Lambda $ varying is more likely than an ensemble where both $\Lambda$ and $\delta$ vary.  The point is that one has to be very cautious about reasoning from the properties of a posited but unobservable ensemble because one can just make things up to fit the data.  Without any independent check on the properties of the ensemble the fact that one can manipulate the assumptions you make about the ensemble to make an outcome seem probably does not in any way constitute evidence for the existence of that ensemble. For example, Garriga and Vilenkin observed\cite{GV} that the argument comes out looking the best if one considers varying a different parameter which is $\Lambda \delta^{3}$.  But this doesn't add any strength to the claim, both because the value of $\delta$ remains unexplained and because, when a false argument has many possible versions, there will always be one that fits the data best.  The flexibility of tuning a false argument to fit the data better does not provide evidence its underlying assumptions are true. 

We can contrast this with the argument made above in the context of cosmological natural selection, where there is an independent argument for $\delta$ to be small.    

Weinberg's prediction was made a decade before the discovery of dark energy, and in science this is not nothing; sometimes a strong intuition can produce a correct prediction even if the logic can be objected to.  But this cannot be used as evidence for the assumption that there really is an ensemble of universes, as the argument from that assumption to the prediction was fallacious, for the reason just explained.  One should also credit Sorkin for correctly predicting the value of the cosmological constant\cite{sorkin}, but for most theorists this doesn't strongly increase their confidence in the causal set theory on which Sorkin's prediction was based. 

These concerns are deepened by the measure problem in eternal inflation.  This arises because there are an infinite number of bubble universes created.  When one has infinite ensembles then assertions of predictions based on relative frequencies become highly problematic.  Any claim that outcome $A$ is more probable than outcome $B$ is problematic when the numbers of $A$ and $B$ are infinite.  The ratios of relative frequencies, 
$N(A)/N(B)$ are then undefined.  

There is a literature whose authors experiment with different measures on these infinite sets which give definitions of the ratios and hence relative frequencies.  The challenge is to avoid various paradoxes, some of which bedevil any application of probability theory to infinite sets, others of which are special to cosmology.  However even if a measure that succeeds in avoiding all the paradoxes were found that would in no way serve to increase the likelihood that the eternal inflation scenario is correct, it would just be another instance of making up the specification of an unobservable ensemble in order to get what one wants from it.  The fact that there may be a best version of a false claim does not increase the credibility of that claim, in the absence of any independent verification of it.  

\subsection{The case of negative cosmological constant in string theory}

A good illustration of these issues is the case of negative cosmological constant in string theory\cite{ellis-smolin}.  There is evidence for a countable infinity of string vacua with negative values of $\Lambda$, accumulating at zero\cite{neglambda}.  This is to be contrasted with the claims that there is a large, but finite number of vacua with small, positive values of $\Lambda$\cite{KKLT}.  Thus there are so far known infinitely more string vacua with small negative $\Lambda$ than with small positive $\Lambda.$  This of course would change were infinitely more string vacua found with positive $\Lambda$.  Nonetheless it is interesting to examine the consequences, assuming this represents the real situation.

Let us first consider the consequences for cosmological natural selection.  Just as there is a positive critical value $\Lambda_0$ which is the upper bound for positive $\Lambda$ for galaxies to form (assuming $\delta$ is fixed to its current value by its involvement in inflation), there will be a negative $\Lambda_0^-$ that any negative value of $\Lambda$ must exceed if galaxies are to form.   Consequently universes will extremize their fitness if $\Lambda$ is in the range $\Lambda_0^- < \Lambda < \Lambda_0$.  

What happens next depends on whether the fitness function depends strongly on the sign of $\Lambda$.  
Let us first assume it does not.   Then, taking into account that  there are infinitely more vacua 
with $\Lambda_0^- < \Lambda < 0$ than with $0 < \Lambda < \Lambda_0$, a randomly chosen universe will be infinitely more likely to be in the range  $\Lambda_0^- < \Lambda < 0$.  Thus,  on the assumption the fitness function does not strongly depend on the sign of $\Lambda$,  the conjunction of the string landscape with cosmological natural selection would predict $\Lambda$ will be slightly negative\cite{ellis-smolin}.  

However, the assumption that the fitness function does not depend on the sign of $\Lambda$ within the range $\Lambda_0^- < \Lambda < \Lambda_0$ can be questioned\footnote{I am grateful to Ben Freivogel for conversations on this issue.}.   The infinite number of string vacua with $\Lambda <0$ described in \cite{neglambda} are not like small continuations in $\Lambda$ of the universes with small positive $\Lambda$. They are supersymmetric and the size of the six extra dimensions are not small, instead they are comparable to the radius of curvature of the four ordinary dimensions given by $R =| \Lambda |^{-\frac{1}{2}}$.  Being supersymmetric there will not be stable atoms with complex chemistry because electrons will always convert to their bosonic partners to reduce the energy of the ground state. And if nine spatial dimensions are large, the physics is very different.  So the  fitness function is going to be sensitive to the sign of $\Lambda$.  However, the argument need not conclude there, as there may be a subset of these negative $\Lambda$ vacua in which supersymmetry is broken spontaneously at lower energies.  It is also possible that there could be brane-worlds related to these solutions on which the usual $3+1$ dimensional physics lives.  All that is needed to restore the argument is that these occur in a finite fraction of the infinite number of negative $\Lambda$ vacua.  

A similar argument can be run in the case of eternal inflation.  Here the argument will rest on whether the anthropic principle can eliminate the negative $\Lambda$ solutions as being unsuited for intelligent life.  One can speculate that intelligent life is impossible when more than three spatial dimensions are large or supersymmetry is unbroken, but it is not clear how convincing a case can be made for this.  In either setting it is far from clear that we know enough to exclude complex structures leading to life or black hole formation.

\section{Conclusions}

The claim that string theory provides a framework for the unification of the laws of physics has long been bedevilled by three issues:

\begin{itemize}

\item{}{\it The lack of a truly background independent formulation,} which for reasons argued above is essential for a quantum theory of gravity.  The progress on $AdS/CFT$ is impressive but does not clinch the case, for three reasons.  First, it not in a cosmological setting, which requires spatially compact conditions with no boundaries or asymptotic regions.  Second, because the sign of $\Lambda$ is wrong.  Third, there is growing evidence that  the $AdS/CFT$ correspondence is a general feature of gravitational theories and is not unique to string theory\cite{AdS-laurent}.  This is especially evident in the reformulation of general relativity under the name of shape dynamics\cite{shape}.  Unfortunately work on background independent approaches to string or $\cal M$ theory remains inconclusive and has become largely neglected in recent years.

\item{}{\it The lack of a complete demonstration of perturbative finiteness. }  My understanding is that there has been some progress on this issue, but it remains unresolved.  

\item{}{\it  The landscape problem. }

\end{itemize} 

In this essay I have argued that the landscape problem as it arises in string theory is symptomatic of a much older and deeper issue: that the completion of a scientific understanding of our universe requires that we not only know what the laws of nature are, but explain why these are the laws.  Thus, what is at stake is whether Leibniz's old dream  that we can give a sufficient reason for every physically meaningful property of the universe can be realized.

I considered here three kinds of cosmological scenarios within which we could search for a scientific solution to the landscape problem.  The results of this analysis were

\begin{enumerate}

\item{} Linear cyclic models could provide a solution  to the landscape problem only if there is an attractor on the landscape corresponding to the standard model and the changes from generation to generation are sufficiently small that the laws can converge to the attractor. However there appears to be no mechanism for the special tunings of the standard model to play a role and hence be explained.

\item{} Branching cosmological models offer the possibility of explaining more.  Branching cyclic models offer an elegant possibility for explaining the cosmological initial conditions without inflation, and they could explain the choice of effective theory under the same conditions as the linear cyclic models.  Cosmological natural selection naturally explains the actual fine tunings of the standard model and is the only scenario to do so.  It also is the only scenario to make falsifiable predictions for present observations.  

\item{} Pluralistic cosmological models such as eternal inflation offer poor prospects for resolving the landscape problem.  Even if there are attractors in the landscape there is no reason for the population to converge to them as there is no reason for the population of universes to be dominated by long chains of descent with incremental changes generation to generation.  Any connection to the real world requires a strong imposition of the anthropic principle which is very unlikely to yield falsifiable predictions.  

\end{enumerate}

String theory brought the landscape issue into focus but, as we have seen, it was inevitable that as physics progressed we would have encountered the problem of explaining how the universe chose its laws. We can call this the generalized landscape problem.   Whether string theory is the right theory of unification or not, it is clear that this general landscape problem must be solved.  But as we have seen, this problem can only be solved  if we abandon the idea that ultimate explanations in physics are to be given in terms of laws organized according to the Newtonian paradigm, with timeless laws acting on a timeless space of states.  

Above all this must apply to whatever theory unifies the different effective theories that make up the landscape.   Within string theory the search for this unification has largely proceeded along traditional lines.

Any solution to the landscape problem must then transcend the Newtonian paradigm.  As Wheeler, Dirac and Pierce understood, laws must evolve to be explained.  It is likely also that the absolute distinction between laws and states must 
break down\footnote{A matrix model which serves as an example of a breakdown of the distinction between law and state is described in \cite{me-breakdown}.}.  Our mandate is then to invent new kinds of theories that answer these challenges, while staying true to the demands that theories make predictions by which they can be falsified. The still open problem of giving string theory or $\cal M$ theory a background independent formulation that would be the setting to resolve the landscape issue should be re-examined in this light.   The main lesson which can be drawn from the successes and failures of attempts to resolve the landscape problem surveyed here is that theories which embrace the evolution of laws  have a better chance to make falsifiable predictions than do theories which try to hold onto to the notion that law is eternal.

\section*{Acknowledgements}

I am grateful to Roberto Mangabeira Unger for proposing a collaboration on evolving laws of nature which stimulated and sharpened many of the ideas and arguments contained here.   Results of our joint work, on which this essay is partially based, will be published in \cite{withroberto}. I am grateful to many colleagues at Perimeter Institute including Niayesh Afshordi,  Latham Bole and Neil Turok for conversations on these issues. I am especially grateful to Matt Johnson for a careful reading of the manuscript and for suggestions which improved it.   I have also learned a great deal from the opportunity to listen to conferences where various attempts to resolve the landscape problem were critically examined.  Research at Perimeter Institute
for Theoretical Physics is supported in part by the Government of
Canada through NSERC and by the Province of Ontario through MRI.


\begin{thebibliography}{99}

\bibitem{LOTC}L. Smolin {\it The Life of the Cosmos}, 
1997 from Oxford University Press (in the USA),
Weidenfeld and Nicolson (in the United Kingdom) and Einaudi Editorici
(in Italy.)

\bibitem{Andyland}A. Strominger, 
  {\it SUPERSTRINGS WITH TORSION,}
  Nucl.\ Phys.\ B {\bf 274}, 253 (1986).
 
  


\bibitem{KKLT}Shamit Kachru, Renata Kallosh, Andrei Linde, Sandip P. Trivedi, 
{\it de Sitter Vacua in String Theory}, hep-th/0301240, Phys.Rev. D68 (2003) 046005. 


\bibitem{lennyland}L. Susskind, {\it   
The Anthropic Landscape of String Theory}, hep-th/0302219;
{\bf The Cosmic Landscape: String Theory and the Illusion of Intelligent Design} (New York: Little Brown, 2006).

\bibitem{Vilenkin-eternali}Alexander Vilenkin, Phys.Rev.D27, 2848 (1983). 

\bibitem{Linde-eternali}A. Linde, { \it    The Inflationary Universe}
Journal-ref: Rept.Prog.Phys. 47 (1984) 925; 
 Andrei Linde, Dmitri Linde, Arthur Mezhlumian, 
{\it  From the Big Bang Theory to the Theory of a Stationary Universe   }, 
gr-qc/9306035,  Phys.Rev. D49 (1994) 1783;
Juan Garcia-Bellido, Andrei Linde, {\it  Stationarity of Inflation and Predictions of Quantum Cosmology  },
 hep-th/9408023, Phys.Rev. D51 (1995) 429; 
 Jaume Garriga, Alexander Vilenkin, {\it  A prescription for probabilities in eternal inflation  }, 
 gr-qc/0102090, Phys.Rev. D64 (2001) 023507.  

\bibitem{PaulNeil}P.~J.~Steinhardt and N.~Turok,
{\it A cyclic model of the universe,'}
Science {\bf 296}, 1436 (2002); 
{\it Cosmic evolution in a cyclic universe,}
Phys.\ Rev.\ D {\bf 65}, 126003 (2002)
[arXiv:hep-th/0111098].

\bibitem{ccc}R. Penrose, {\bf Cycles of Time},  Random House, 2011. 

\bibitem{roger-circles}V. G. Gurzadyan, R. Penrose , {\it   CCC-predicted low-variance circles in CMB sky and LCDM  }.  arXiv:1104.5675; {\it  More on the low variance circles in CMB sky}, arXiv:1012.1486{\it  Concentric circles in WMAP data may provide evidence of violent pre-Big-Bang activity}, 
arXiv:1011.3706. 



\bibitem{nocircles}I. K. Wehus, H. K. Eriksen, {\it A search for concentric circles in the 7-year WMAP temperature sky maps,} 
 arXiv:1012.1268, ApJ, 733, L29 (2011); Adam Moss, Douglas Scott, James P. Zibin, 
 {\it No evidence for anomalously low variance circles on the sky}, arXiv:1012.1305;
 Amir Hajian, {\it Are There Echoes From The Pre-Big Bang Universe? A Search for Low Variance Circles in the CMB Sky}, 
 arXiv:1012.1656.  

\bibitem{phoenix}Jean-Luc Lehners,  {\it  Diversity in the Phoenix Universe},  arXiv:1107.4551;
Jean-Luc Lehners, Paul J. Steinhardt, Neil Turok, {\it The Return of the Phoenix Universe},    arXiv:0910.0834,  Int.J.Mod.Phys.D18:2231-2235,2009.


\bibitem{CNS}L. Smolin, 1992a.   "Did the Universe Evolve?" 
Class. Quantum Grav.  9,  173-191.

\bibitem{CNS-2}L. Smolin, {\it On the fate of black hole singularities 
and the parameters of the standard model}, 
 gr-qc/9404011, CGPG-94/3-5 ;  {\it Using neutrons stars and primordial black holes to
test theories of quantum gravity}, astro-ph/9712189; 
L. Smolin, {\it Cosmology as a problem in 
critical phenomena}  
in the proceedings of the Guanajuato Conference
on Complex systems and binary networks, (Springer,1995),
eds. R. Lopez-Pena, R. Capovilla, R. Garcia-Pelayo,
H. Waalebroeck and F. Zertuche.
gr-qc/9505022; {\it Experimental Signatures of Quantum Gravity}  
in the Proceedings of the Fourth Drexel Conference on Quantum
Nonintegrability, International Press, to appear,
gr-qc/9503027.


\bibitem{me-alt-AP}L. Smolin,  {\it Scientific alternatives to the anthropic principle,'},  arXiv:hep-th/0407213,
Contribution to "Universe or Multiverse", ed. by Bernard Carr et. al.,  published by Cambridge University Press.

\bibitem{mestatus}L. Smolin, {\it The status of cosmological natural selection}, arXiv:hep-th/0612185.  

\bibitem{hybrids}M. Johnson and J-L Lehners,   {\it Cycles in the Multiverse} ,  arXiv:1112.3360. 

\bibitem{popper}Karl Popper, {\it Conjectures and Refutations}, 
London: Routledge and Keagan Paul, 1963, pp. 33-39; from Theodore 
Schick, ed., Readings in the Philosophy of Science, Mountain View, CA: Mayfield Publishing Company, 2000, pp. 9-13; {\it The open society and its enemies} (1946); 
{\it The logic of scientific discovery} (1959). 

\bibitem{wheeler}J. A. Wheeler, in {\it Black holes, gravitational
waves and cosmology} eds. Martin Rees, Remo Ruffini and J. A. 
Wheeler, New York: Gordon and Breach, 1974.

\bibitem{bounce-old}See, for example, V. P. Frolov,   M. A. Markov and M. A. Mukhanov, 1989.  
"Through a black hole into a new universe?",  Phys. Lett. B216 272-276;
A. Lawrence and E. Martinec, 1996.  "String field theory in curved 
space-time and the resolution of spacelike singularities"  Class. 
and Quant. Grav. 13, 63;  hep-th/9509149;
E. Martinec, 1995.  "Spacelike singularities in string theory".  
Class. and Quant. Grav. 12, 941-950.   hep-th/9412074.

\bibitem{LQC-bounce}Martin Bojowald,''Isotropic Loop Quantum Cosmology'',
Class.Quant.Grav. 19 (2002) 2717-2742,   gr-qc/0202077;
    ``Inflation from Quantum Geometry'', gr-qc/0206054;
    ``The Semiclassical Limit of Loop Quantum Cosmology'',
gr-qc/0105113, Class.Quant.Grav. 18 (2001) L109-L116;
`` Dynamical Initial Conditions in Quantum Cosmology'',
gr-qc/0104072,  Phys.Rev.Lett. 87 (2001) 121301; 
S. Tsujikawa, P.  Singh, R. Maartens ,{\it Loop quantum gravity effects on inflation and the CMB}, astro-ph/0311015 

\bibitem{jr-bounce}Rodolfo Gambini, Jorge Pullin,
{\it   Discrete quantum gravity: a mechanism for selecting the value of fundamental constants  }, 
gr-qc/0306095, Int.J.Mod.Phys. D12 (2003) 1775-1782. 

\bibitem{dirac}Paul A M Dirac, {\it The relation between mathematics and physics}, The Proceedings of the Royal Society (Edinburgh)  59 (1939) 122-129.

\bibitem{CSP}Charles Sanders Peirce,  {\it The architecture of theories}, The Monist, 1891, reprinted in {\it Philosophical Writings of Peirce}, ed. J. Buchler. New York: Dover, 1955.

\bibitem{leibniz}Gottfried Wilhelm Leibniz, THE MONADOLOGY, 1698, translated by Robert Latta, availabe at http://oregonstate.edu/instruct/phl302/texts/leibniz/monadology.html; Leibniz, Gottfried Wilhelm (Oxford Philosophical Texts), ed. by Francks, Richard and Woolhouse, R. S., Oxford Univ. Press, 1999; H. G. Alexander, ed., The Leibniz- Clarke Correspondence, Manchester University Press, 1956, for an annotated selection, see http://www.bun.kyoto-u.ac.jp?suchii/leibniz-clarke.html.

\bibitem{Leibniz-psfr}Gottfried Wilhelm Leibniz, , {\it Philosophical writings}, English translation, Mary Morris. London: Everyman's library, no. 905, pages 193-194.

\bibitem{nongauss}R. Holman, Andrew J. Tolley, {\it Enhanced Non-Gaussianity from Excited Initial States}, arXiv:0710.1302, JCAP 0805:001,2008.

\bibitem{recycling}Jaume Garriga, Alexander Vilenkin, {\it Recycling universe}, arXiv:astro-ph/9707292, Phys.Rev. D57 (1998) 2230-2244.  

\bibitem{infinitecollisions}Alan  Guth, Phys. Rev. D23, 876 and Nucl. Phys. B212, 321 

\bibitem{critiques}
T. Rothman and G.F.R. Ellis, 1993.  "Smolin's natural 
selection hypothesis", Q. J. R. astr. Soc. 34, 201-212; A. Vilenkin, {\it On cosmic natural selection}, hep-th/0610051;  E. R. Harrison,  {\it THE NATURAL SELECTION OF 
UNIVERSES CONTAINING INTELLIGENT LIFE}, 
R.A.S. QUARTERLY JOURNAL V. 36, NO. 3/SEP, P. 193, 1995;  J. Silk, Science, 227 (1997) 644.

\bibitem{bethe-brown}
G. E. Brown and H. A. Bethe, Astro. J. 423 
(1994) 659; 436 (1994) 843, G. E. Brown, Nucl. Phys. A574 
(1994) 217; G. E. Brown, ``Kaon condensation in dense matter";
H. A. Bethe and G. E. Brown, ``Observational constraints on
the maximum neutron star mass", preprints.

\bibitem{LP}James M. Lattimer, M. Prakash, 
{\it What a Two Solar Mass Neutron Star Really Means}. 
4. arXiv:1012.3208 , to appear in Gerry Brown's Festschrift; Editor: Sabine Lee (World Scientific) 

\bibitem{bubbles}Stephen M. Feeney, Matthew C. Johnson, Daniel J. Mortlock, Hiranya V. Peiris
{\it    First Observational Tests of Eternal Inflation: Analysis Methods and WMAP 7-Year Results  },
arXiv:1012.3667,  Phys.Rev.D84:043507,2011; 
Anthony Aguirre, Matthew C. Johnson, {\it   A status report on the observability of cosmic bubble collisions }, arXiv:0908.4105,  Rept.Prog.Phys.74:074901,2011 


\bibitem{AP}
B. Carter, 1974.  in  
Confrontation of Cosmological Theories with Observational Data,
IAU Symposium No. 63, ed. M. Longair Dordrecht: Reidel.  p. 291; B. J. Carr and M. J. Rees, 1979.  Nature 278, 605;
B. Carter, 1967.  "The significance of numerical coincidences in nature", 
unpublished preprint, Cambridge University; 
J. D. Barrow and F. J. Tipler  
{\it The Anthropic Cosmological Principle}  
(Oxford University Press,Oxford,1986).

\bibitem{TTWP}L. Smolin, {\bf The Trouble with Physics}, Houghton-Mifflin, Boston, 2006.  





\bibitem{weinberg}S. Weinberg, Phys. Rev. Lett 59, 2067 (1987); {\it 
A Priori Probability Distribution of the Cosmological Constant}. 
Phys.Rev. D61 (2000) 103505,     astro-ph/0002387; 
{\it The Cosmological Constant Problems}, 
astro-ph/0005265. 




\bibitem{varymore}M.J. Rees, Complexity 3 17 (1997); in Fred HoyleÕs Universe, eds. C. Wickramasinghe et al.
(Kluwer, Dordrecht, 2003) [astro-ph/0401424]; M. Tegmark and M.J. Rees, Astrophys. J. 499, 526 (1998) [astro-ph/9709058]; Michael L. Graesser, Stephen D.H. Hsu, Alejandro Jenkins, Mark B. Wise, 
{\it    Anthropic Distribution for Cosmological Constant and Primordial Density Perturbations   },
hep-th/0407174. 


\bibitem{GV}Jaume Garriga, Alexander Vilenkin, {\it  Anthropic prediction for Lambda and the Q catastrophe,}
arXiv:hep-th/0508005,  Prog.Theor.Phys.Suppl. 163 (2006) 245-257. 


\bibitem{kragh}Helge Kragh, {\bf Higher Speculations: grand theories and failed revolutions in physics and cosmology},
Oxford University Press 2011.


\bibitem{sorkin}Maqbool Ahmed, Scott Dodelson, 
Patrick B. Greene, Rafael Sorkin, {\it Everpresent Lambda},
astro-ph/0209274.

\bibitem{ellis-smolin}George Ellis and Lee Smolin {\it The weak anthropic principle and the landscape of string theory}   arXiv:hep-th:0901.2414

\bibitem{neglambda} O DeWolfe, A Giryavets, S Kachru and W Taylor. ÒType IIA Moduli Stabiliza- tionÓ. arXiv: hep-th/0505160 (2005);  J Shelton, W Taylor, and B Wecht, ÒGeneralized Flux VacuaÓ. arXiv: hep- th/0607015 (2006).

\bibitem{AdS-laurent} Laurent Freidel,  {\it  Reconstructing AdS/CFT   },  arXiv:0804.0632. 

\bibitem{shape}Henrique Gomes, Sean Gryb, Tim Koslowski, Flavio Mercati, {\it  The gravity/CFT correspondence     }  arXiv:1105.0938. 

\bibitem{withroberto}Roberto Mangabeira Unger and Lee Smolin, book manuscript in preparation.  See also, 
http://pirsa.org/10050053/,   http://pirsa.org/08100049/.  

\bibitem{me-breakdown}L. Smolin, {\it A unification of the state with the dynamical law}, preprint in preparation, see also
http://pirsa.org/11100113/.  

\end{thebibliography}
\end{document}